\documentclass{aastex63}

\received{\today}
\revised{\today}
\accepted{\today}
\submitjournal{ApJ}

\shorttitle{Origin of Strong Linear Polarization}
\shortauthors{Wang et al.}
\graphicspath{{./}{figures/}}

\begin{document}


\title{Origin of Strong Linear Polarization from Fast Radio Bursts}

\author{Runna Wang}
\affiliation{%
 School of Physical Science and Technology, Southwest Jiaotong University, Chengdu 610031,China
}%
\author[0000-0003-1039-9521]{Siming Liu}%
\affiliation{Key laboratory of dark matter and space astronomy, Purple Mountain Observatory, Chinese Academy of Sciences, Nanjing 210023, People's Republic of China
}%
\affiliation{%
 School of Physical Science and Technology, Southwest Jiaotong University, Chengdu 610031,China
 }

\author{Anda Xiong}
\affiliation{
 School of Physics and Astronomy, University of Birmingham, Birmingham, United Kingdom\\
}%
\author{Qi-Hui Chen}
\affiliation{%
 School of Physical Science and Technology, Southwest Jiaotong University, Chengdu 610031,China
 }
 \author{Fengrong Zhu}
\affiliation{%
 School of Physical Science and Technology, Southwest Jiaotong University, Chengdu 610031,China
}%


\date{\today}

\correspondingauthor{Siming Liu}
\email{liusm@pmo.ac.cn}
\correspondingauthor{Fengrong Zhu}
\email{zhufr@swjtu.edu.cn}




\begin{abstract}

The detection of almost 100\% linearly polarized emission from the fast radio burst source FRB 121102 implies coherent emission of relativistic electrons moving in perpendicular to the ambient magnetic field. The origin of such a particle distribution is very intriguing. Given that FRB 121102 is likely driven by a neutron star, we explored orbits of charged particles trapped in a dipole magnetic field (the St\"{o}rmer problem). Most previous studies focused on particles with relatively low energies so that the guiding center approximation may be applied. High energy particles usually have chaotic orbits except those on a periodic orbit or near stable periodic orbits. Via evaluation of the maximum Lyapunov exponent of orbits of particles launched from the equatorial plane with an axial velocity (the angular velocity sets the length and energy scales of the system), we found  prominent regions of quasi-periodic orbits around stable periodic orbits in the equatorial plane at high-energies. Particles in these orbits oscillate around the equatorial plane and their radial distance from the dipole can vary by a factor of $\sim 2$. Relativistic electrons in such orbits may be responsible for the almost 100\% polarized emission from FRB 121102.

\end{abstract}



\section{Introduction} \label{sec:intro}

The discovery of repeated fast radio burst (FRB) source FRB121102 and its intriguing emission properties have energized the study of FRBs dramatically \citep{2018Natur.553..182M}. The almost 100\% linearly polarized emission detected from this source implies a strongly magnetized environment, and the classical emission theory suggests coherent emission of electrons moving in perpendicular to the ambient magnetic field. There is no evident mechanism that can lead to such an electron distribution. Considering the effective radiative energy loss and propagation effects of high-energy electrons, a beaming distribution is expected.
The fact that this source can produce FRBs repeatedly shows that it should be powered by a compact object, e.g., a neutron star. As a leading order approximation, the magnetic field of a neutron star may be approximated by a magnetic dipole. Motions of a charged particle in a dipole magnetic field have been studied extensively since \citet{1930ZA......1..237S} first reduced this problem to a 2 dimensional Hamiltonian system.

It is well-known that there are quasi-periodic orbits around stable periodic orbits in Hamiltonian systems with 2 degrees of freedom \citep{1990ComPh...4..549S}. For the axis symmetry of the system, motions of a charged particle in a dipole magnetic field can be reduced to a 4 dimensional (4D) Hamiltonian system in the phase space \citep{1965RvGSP...3..255D}, and periodic orbits in the equatorial and Meridian planes have been explored extensively \citep{devogelaere1950, 1975Ap&SS..32..115M,1978CeMec..17..215M, 1990CeMDA..49..327J}. These periodic orbits appear to have a negligible measure in the phase space and are difficult to realize in nature. Quasi-periodic orbits, on the other hand, may have a finite volume in the 4D phase space and be readily detectable \citep{1963RuMaS..18....9A}.

Lyapunov exponents measure the dependence of a trajectory on its initial conditions. A chaotic orbit has at least one positive Lyapunov exponent \citep{doi:10.1142/S0218127400000177}. The maximum Lyapunov exponent of quasi-periodic orbits, on the other hand, should be zero \citep{1963RuMaS..18....9A}. To produce coherent emission, charged particles need to move in synchrony. Chaotic motions clearly do not lead to coherent emission. Electrons on quasi-periodic motions however are expected to give rise to coherent emission. Via calculation of the Lyapunov exponents of orbits of charged particles trapped in a dipole magnetic field, we scanned the phase space of particles launched from the equatorial plane with a vanishing radial velocity. Besides the low energy regime that has been studied extensively with the guiding center approximation \citep{1963RvGSP...1..283N}, we also found high energy regimes of quasi-periodic orbits associated with stable periodic orbits in the equatorial plane \citep{devogelaere1950, 1965RvGSP...3..255D}. The areas of these regimes in the phase plane of the initial conditions are comparable to that of the low energy regime, and particles in these orbits may produce coherent emission responsible for strongly linearly polarized emission from FRBs.

This letter is organized as the following. In section \ref{stormer}, we briefly summarize the St\"{o}rmer problem. Section \ref{results} shows the key results of this study. Conclusions are drawn in section \ref{con}.

\section{The St{\"o}rmer Problem}
\label{stormer}

\citet{1930ZA......1..237S} first carried out a systemic study of motion of charged particles in a dipole magnetic field. He found that considering the axis symmetry of the system, the problem can be reduced to a 2D Hamiltonian one with an effective potential in the dimensionless formula. For the sake of completeness, we follow \citet{1965RvGSP...3..255D} and summarize this problem briefly here.

Considering a particle with an electric charge $q$ and rest mass $m$ in a dipole magnetic field ${\bf B}$ with the dipole pointing to the $z$ direction at the origin. We will adopt a spherical coordinate with $r$ the radial distance of a point to the origin and $\rho$ its projection on the equatorial plane ($x$, $y$). Then we have $r^2=\rho^2+z^2$. The vertical and azimuthal coordinates are indicated by $z$ and $\phi$, respectively. The magnetic field line is then given by $r = r_0\sin^2\theta$, where $\theta$ is the polar angle. The relativistic Hamiltonian $H_R$ of the system is given by
\begin{equation}
H_R=\left\{m^2c^4+c^2\left[p_z^2+p_\rho^2+\left(\frac{p_\phi}{\rho}-qA_\phi\right)^2\right]\right\}^\frac{1}{2},
\label{eq:Hamiltonian-1}
\end{equation}
where
\begin{eqnarray}
p_z&=&\gamma m\dot{z},\\
p_\rho&=&\gamma m\dot{\rho},\\
p_\phi&=&\gamma m\rho^2\dot{\phi}+q\rho A_\phi,
\end{eqnarray}
where an upper dot indicates a derivative with respect to time $t$, $\gamma=(1-{v^2}/{c^2})^{-{1}/{2}}$ is the Lorentz factor of the particle moving with a speed of $v$, and $c$ is the speed of light.
$A_\phi=M\rho/r^3$ is the amplitude of the vector potential ${\bf A}$, and $M$ is the magnetic moment of the dipole.

Since the Hamiltonian is independent of $t$, the energy is a constant of motion
\begin{equation}
H_R=\gamma mc^2={\rm constant},
\label{eq:Hamiltonian-1-1}
\end{equation}
and, hence, $\gamma$ is a constant too. The constancy of $H_R$ and $\gamma$ make it possible to introduce a new Hamiltonian, $H$, which is equivalent to $H_R$ and has a simpler form that resembles a non-relativistic Hamiltonian:
\begin{equation}
H=\frac{1}{2\gamma m}\left[p_z^2+p_\rho^2+(\frac{p_\phi}{\rho}-qA_\phi)^2\right].
\label{eq:Hamiltonian-2}
\end{equation}
And $H$ is also conserved
\begin{equation}
H=\frac{1}{2}\gamma mv^2={\rm constant}.
\label{eq:Hamiltonian0}
\end{equation}
For the sake of simplicity, we choose $H$ instead of $H_R$ in this study. 

Considering the conservation of $p_\phi$, we have a characteristic length scale $L= qM/p_\phi$. The corresponding energy and momentum scale are $p_\phi^2/(\gamma mL^2)= p_\phi^4/(\gamma mq^2M^2)$, $p_\phi/L=p_\phi^2/(qM)$, respectively.
Then we have the dimensionless Hamiltonian:
\begin{equation}
h=\frac{1}{2}\left(p_z^2+p_\rho^2\right)+\frac{1}{2}\left(\frac{1}{\rho}-\frac{\rho}{r^3}\right)^2\,.
\label{eq:Hamiltonian}
\end{equation}
One should note that the effective potential
\begin{equation}
    V = \frac{1}{2}\left(\frac{1}{\rho}-\frac{\rho}{r^3}\right)^2
\end{equation}
is proportional to $(\gamma m\rho\dot{\phi})^2$ and has a saddle point at $\rho=r=2$, where $V=1/32$ and particles make circular motion around the dipole in the equatorial plane with the centrifugal force balanced by the Lorentz force. Particles injected within $\rho= 2$ and with an energy $h<1/32$ will be trapped. The energy contours are tangential to the $z$ axis at the origin so that the polar angle $\theta$ approaches 0 as particles moving toward the origin.

The Lorentz factor of a particle $\gamma$ can be expressed with the dimensionless energy $h$ and $p_\phi$ as:
\begin{equation}
    \gamma=\left[1+{2hp_\phi^4\over c^2m^2q^2M^2}\right]^{1/2}\,.
\label{lorentz1}
\end{equation}
At the saddle point, trapped particles reach the maximum energy: $p_\phi = 2\rho q A_\phi = 2 q M/\rho$. Then the maximum Lorentz factor in trapped orbits is given by
\begin{equation}
    \gamma_{\rm max}=\left[1+{q^2M^2\over c^2m^2\rho^4}\right]^{1/2}\,.
    \label{lorentzm}
\end{equation}

The Hamiltonian equations are given by:
\begin{eqnarray}
\dot{z}&=&p_z,\\
\dot{\rho}&=&p_\rho,\\
\dot{p_z}&=&\frac{3z\rho}{(z^2+\rho^2)^{\frac{5}{2}}}\left[\frac{\rho}{(z^2+\rho^2)^{\frac{3}{2}}}-\frac{1}{\rho}\right],\\
\dot{p_\rho}&=&\left[\frac{1}{\rho^2}+\frac{z^2-2\rho^2}{(z^2+\rho^2)^{\frac{5}{2}}}\right]\left[\frac{1}{\rho}-\frac{\rho}{(z^2+\rho^2)^{\frac{3}{2}}}\right].
\label{eq:Differential equation}
\end{eqnarray}
For given initial conditions of $\rho_0$, $z_0$, $p_{\rho0}$, $p_{z0}$, these equations can be solved numerically to obtain the trajectory of the particle in the phase space and the Lyapunov exponents can be evaluated accordingly \citep{1985PhyD...16..285W}.

\section{Results}
\label{results}

The Lyapunov exponents are a measure of orbital sensitivity to the initial conditions, and represent the average exponential convergence or divergence rate between adjacent orbits in the phase space. An n-dimensional dynamic system has n Lyapunov exponents, whereas the largest of which is called the largest Lyapunov exponent (LLE). A positive LLE usually indicates chaos. In our case, each set of initial values lead to a group of four Lyapunov exponents. If there is one positive Lyapunov exponent then the system is considered chaotic, whereas two positive Lyapunov exponents means hyper chaotic. If there is no positive Lyapunov exponent, then the orbits are stable and quasi-periodic \citep{doi:10.1142/S0218127400000177}.

\begin{figure}[htp]
\includegraphics[width=0.45\textwidth]{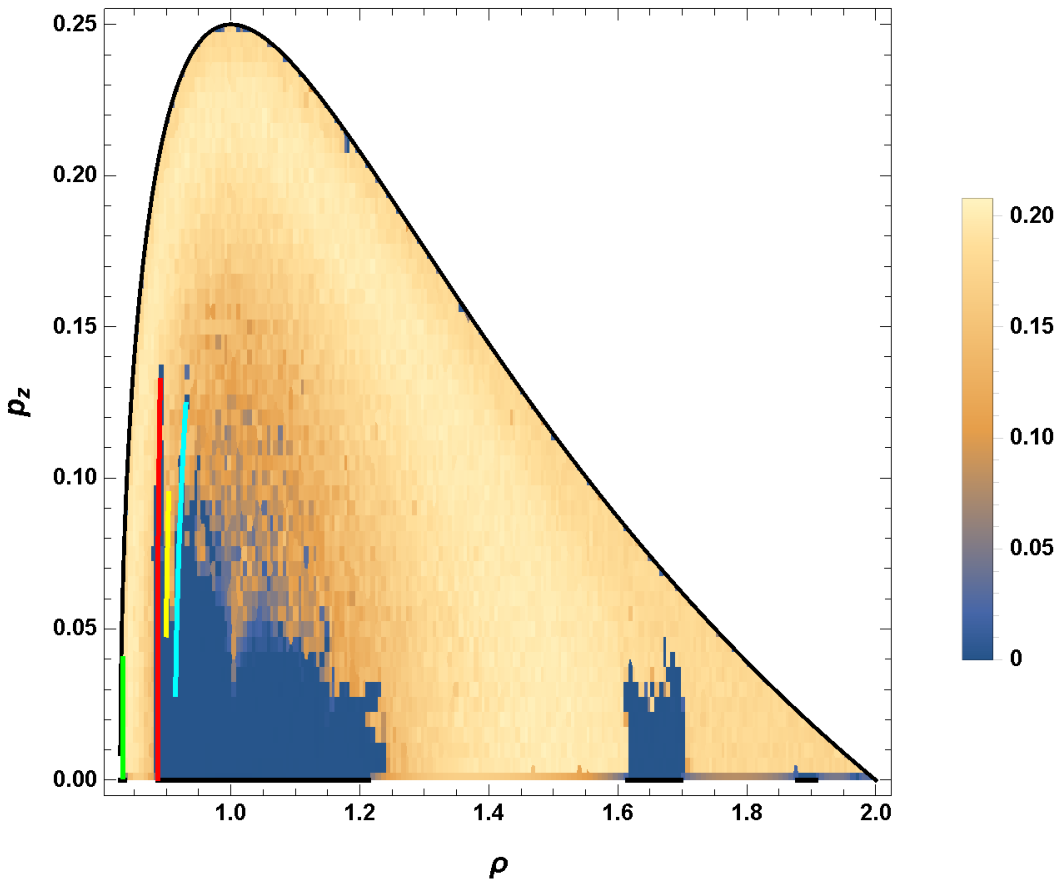}
\includegraphics[width=0.50\textwidth]{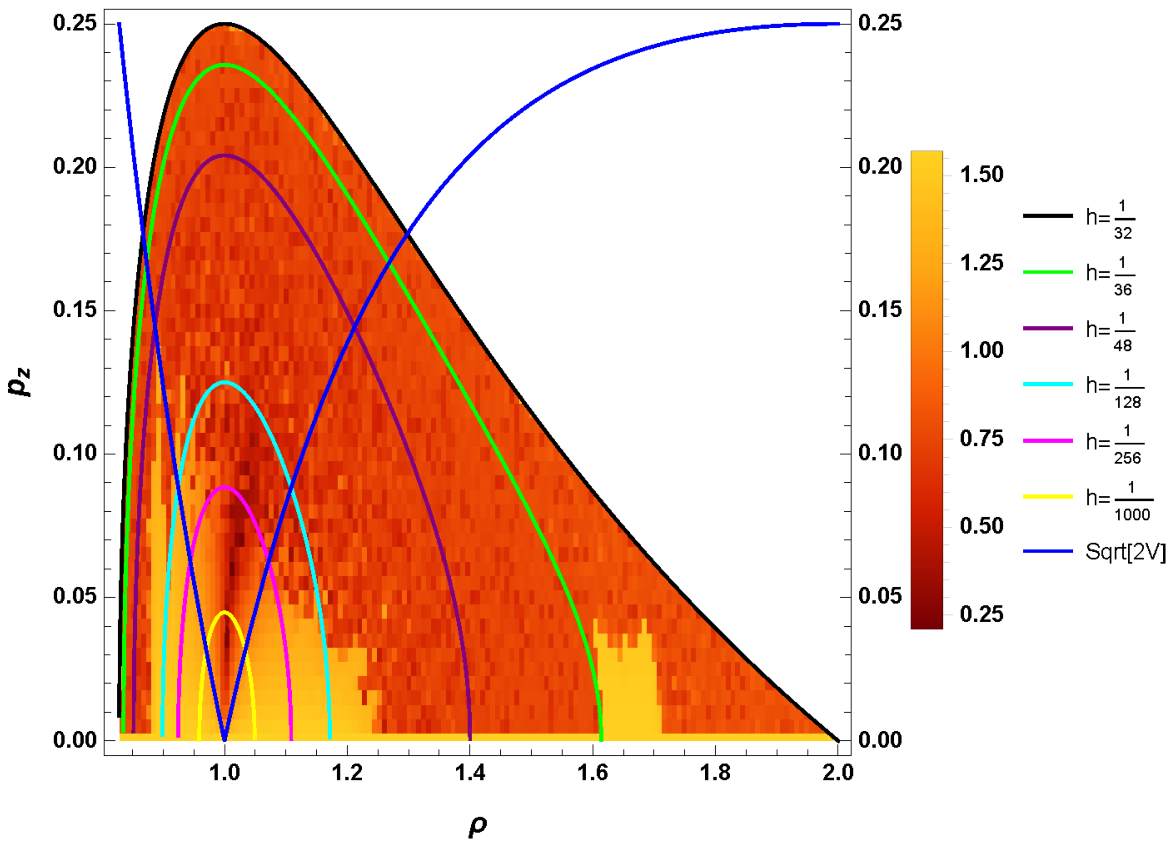}
\caption{\label{fig1} Left: distribution of the LLE in the ($\rho$, $p_z$) plane with $z=0$ and $p_\rho=0$. The line segments indicate stable periodic orbits. The outline is the energy contour of $1/32$. The energies of other energy contours are indicated in the figure. Right: distribution of the minimum polar angle for a duration of 1000. The blue line shows the perpendicular momentum in the equatorial plane.}
\end{figure}

We scanned the initial phase plane, where particles are injected from the equatorial plane with a vanishing radial velocity and calculated the Lyapunov exponents of orbit for each set of initial conditions.
The left panel shows the distribution of the LLE in the ($\rho$, $p_z$) plane.
The outer line of the colored region corresponds to the energy of $1/32$ for the boundary of trapped orbits. Most other orbits outside this region will extend to infinity. There are two evident blue regions of quasi-periodic orbits. The one close to the minimum of $V$, i.e. $\rho=1$, has lower energies and has been studied extensively with the guiding center approximation \citep{1963RvGSP...1..283N}. There is also a set of quasi-periodic orbits with higher energies near $\rho=1.65$. In this figure, we also use black (along the $\rho$ axis) and colored line segments to indicate stable periodic orbits in the equatorial and Meridian planes, respectively \citep{devogelaere1950, 1978CeMec..17..215M}.
The blue regions are associated with stable periodic orbits. The boundary between the blue and orange areas, equivalently as the boundary between quasi-periodic and chaotic dynamical state, is observed with fractality. This complexity of fractal boundary is naturally intrinsically associated to chaos. These quasi-periodicity results from the KAM stability \citep{1963RuMaS..18....9A}.

\begin{figure}[htb]
\centering
\includegraphics[width=0.45\textwidth]{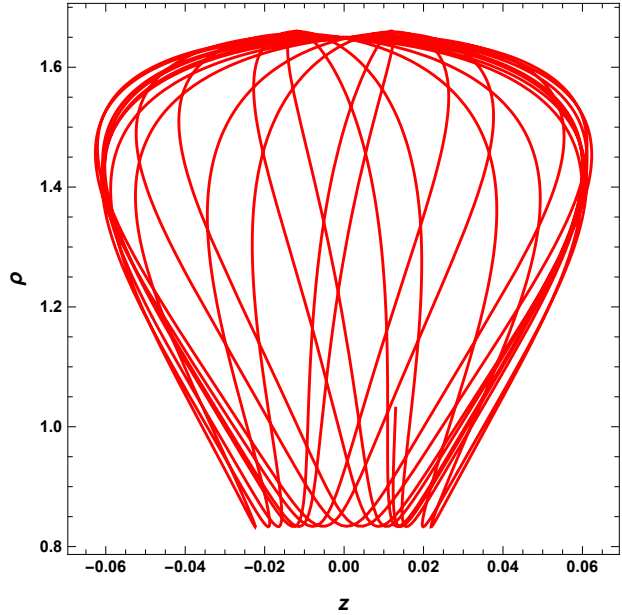}
\caption{\label{fig2} 
Typical orbit (for a duration of 250) of a particle launched from the high-energy region of quasi-periodic orbits discovered in this paper. The initial conditions are $\rho_0=1.65$, $z_0=0$, $p_{\rho0} = 0$, $p_{z0}=0.02$.}
\end{figure}

The right panel of Figure \ref{fig1} shows the distribution of the minimum polar angle of these orbits for a duration of 1000. We note that the red region should be interpreted as upper limits since their values decrease with the increase of the calculation duration. In the orange regions, the minimum polar angle does not depend on the calculation duration. Particles in quasi-periodic orbits seem to always oscillate around the equatorial plane except those near the minimum of $V$ with $\rho=1$. Figure \ref{fig2} shows the trajectory of a particle in the high-energy region of quasi-periodic orbits. It has access to a large radial range so that the guiding center approximation is invalid. Its $z$ coordinate is always more than 20 times smaller than the $\rho$ coordinate, implying motion nearly perpendicular to the magnetic field. For particles to reach the north and/or south poles along magnetic field lines from the radiation belt, one needs a polar angle smaller than $\sim 30^\circ$ \citep{2014Natur.515..531B}. Figure \ref{fig1} shows that most orbits of these particles cannot be quasi-periodic. Therefore the guiding center approximation is not valid for these particles. The guiding center approximation is only valid in a very narrow region near $\rho=1$ as indicated by the shape red feature surrounded by orange zone of quasi-periodic orbits in the right panel of Figure \ref{fig1}.

\begin{figure}[htb]
\centering
\includegraphics[width=0.5\textwidth]{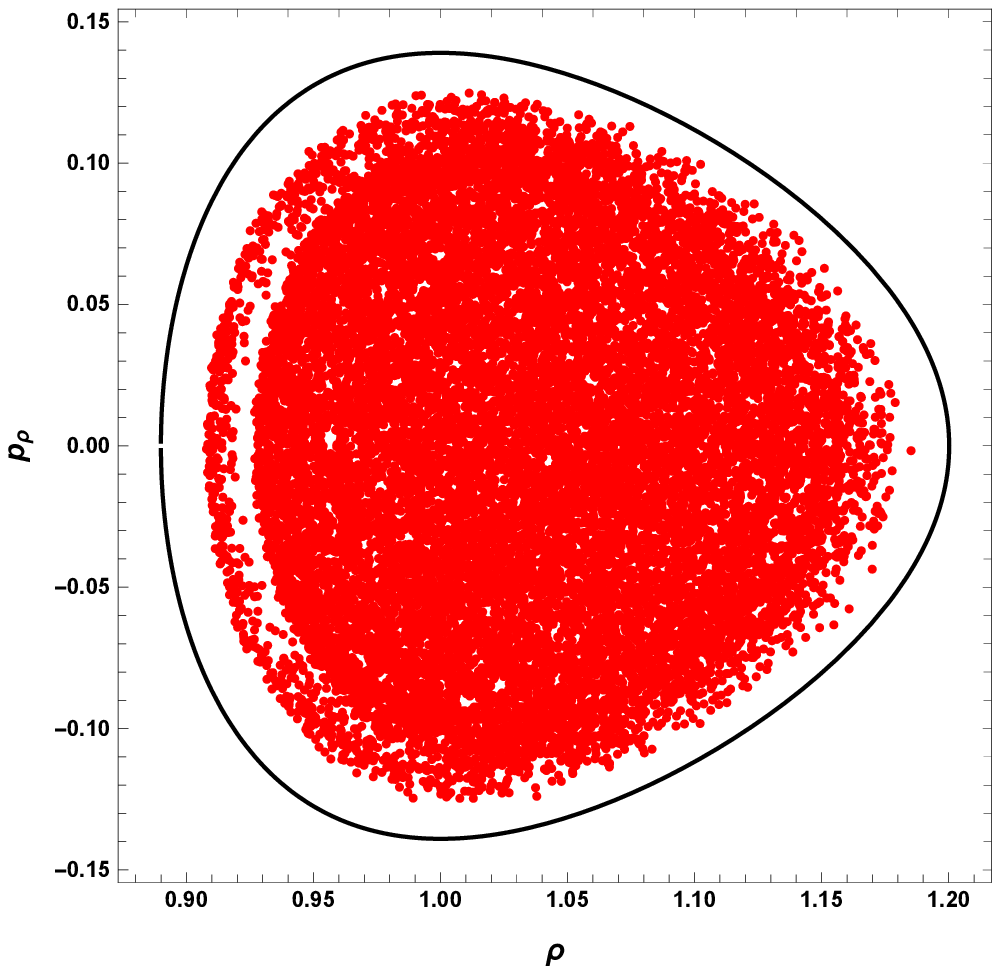}
\includegraphics[width=0.48\textwidth]{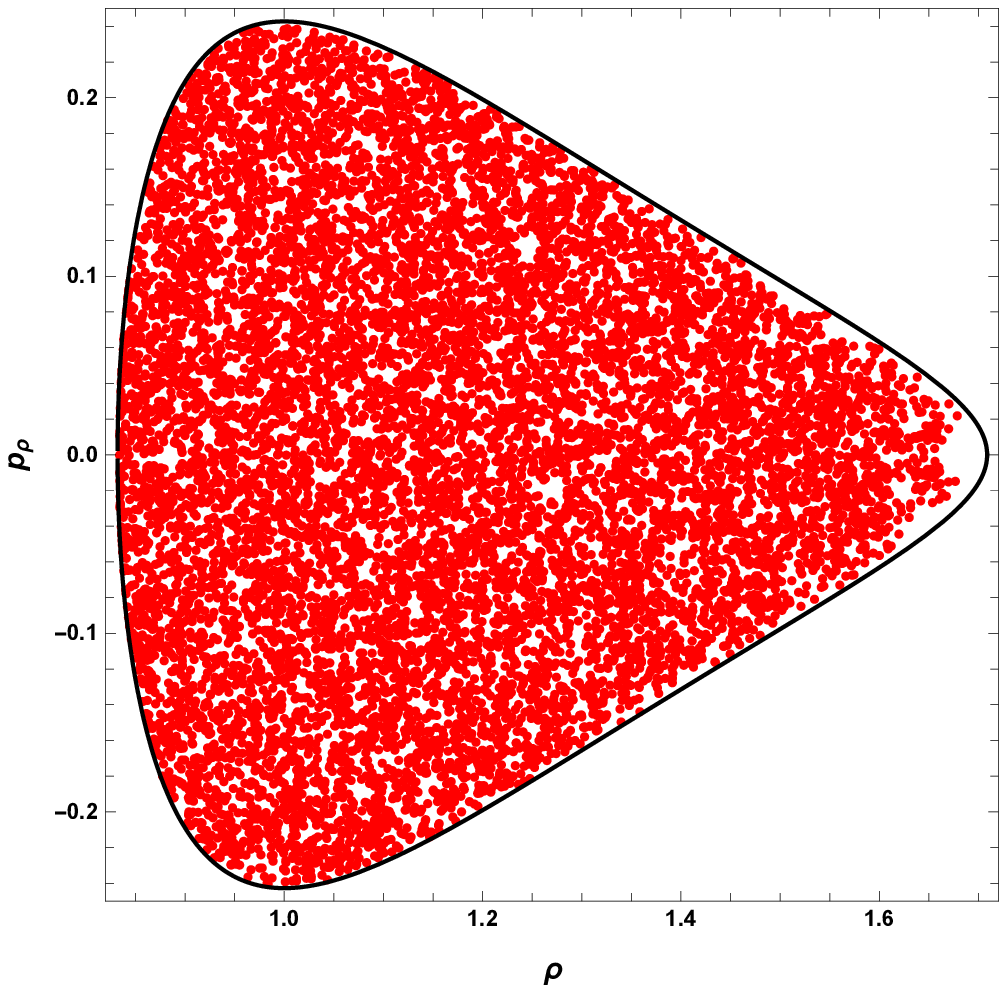}
\caption{\label{fig3} Poincar\'{e} diagram of a chaotic (left) [hyper chaotic (right)] orbit with $\rho_0=0.94565$, $z_0=0$, $p_{\rho0} = 0$, $p_{z0}=0.125 [0.235]$ with a dimensionless energy of 0.00965942 [0.0294594] indicated with the outer contour.}
\end{figure}

Figure \ref{fig3} shows the Poincar\'{e} diagram, that is the points in the ($\rho$, $p_\rho$) plane whenever the particle crosses the equatorial plane upward, of a chaotic (left with one positive Lyapunov exponent) and a hyper chaotic orbit (right with two positive Lyapunov exponents). For the hyper chaotic case, the particle can be arbitrarily close to any points allowed by energy conservation in the phase space (except for an empty region to right occupied by quasi-periodic orbits, see the following discussion). A significant fraction of the phase space allowed by energy conservation is still not reachable for the chaotic orbit. Moreover, there is a vacuum region within the Poincar\'{e} map of the chaotic orbit. We have verified that at the same energy, the point with $\rho_0=0.922$, $z_0=0$, $p_{\rho0} = 0$, $p_{z0}=0.104402$ gives a quasi-periodic orbit. This shows that quasi-periodic orbits occupy a 4D volume in the phase space so that chaotic orbits cannot reach it.

For a given physical length scale $R$ that for simplicity may be associated with the radius of circular motion at the saddle point $2L$, then $p_\phi = 2qM/R$, one can convert each region of quasi-periodic orbits in the 4D phase space into particle distribution in physical space. Of course, all these trapped particles need to be confined within a sphere with a radius of $R$. Since the dimensionless angular momentum $\gamma m\rho\dot{\phi}L/p_\phi$ is given by $(2V)^{1/2}$, in the right panel of Figure \ref{fig1} one can readily get the initial pitch angle of the injected particle. It can be seen that except at the minimum of $V$, where the particle is at rest, the pitch angle of particle with a quasi-periodic orbit is usually greater than $45^\circ$. We therefore expect that particles in quasi-periodic orbits have a momentum distribution centered around $p_\perp = 4[2V(\rho, z)]^{1/2}qM/R^2$ and $p_{||} = 0$, where $p_\perp$ and $p_{||}$ are the perpendicular and parallel momentum, respectively. Coherent emission is expected from such kind of distribution.

For neutron stars, assuming a magnetic field $B$ of $10^{12}$ G at a radius of $10^6$ cm, if the emitting electrons are injected at $10^{10}$ cm, the corresponding magnetic field is about $1$ G,
$\gamma_{\rm max} \simeq 6\times 10^{6}$ for electrons.
(Since no short periodic signal has been detected from repeated FRBs, one may ignore the constraint set by the light cylinder of fast spinning neutron star.)
The corresponding energy loss time via curvature radiation is on the order of $10^{2}$ s, which is much longer than the gyro-period of $\sim 2$ s. However, the emission efficiency is much higher for coherent processes, the quasi-periodic motion may be destroyed on the millisecond scale of FRBs.

On the other hand, if the emitting electrons are injected at $10^9$ cm, the corresponding magnetic field is about $10^3$ G, $\gamma_{\rm max} \simeq 6\times 10^{8}$. The radiative energy loss time will be on the order of $10^{-6}$s, which is much shorter than the duration of FRBs and the gyro-period of $0.2$ s. The electron energy will be lost mostly in the $\gamma$-ray band. The high-energy region of quasi-periodic orbits will be destroyed when one considers the energy loss of relativistic electrons in this case. It is therefore more likely that FRBs are mainly produced by electrons in the low-energy region of quasi-periodic orbits, then the energy of electrons can be much lower. If one attributes the duration of about 1 ms for FRBs to radiative energy loss, the Lorentz factor of the emitting electrons should be about $10^{6}(B/10^3{\rm G})^{-2}\ll\gamma_{\rm max}$. For coherent emission, the energy of electron can be even lower. For electrons in such low energy orbits, the magnetic field is about 8 times higher, the corresponding gyro-period is about $0.3(B/10^3{\rm G})^{-3}$ ms, which is much shorter than the duration of FRBs. Coherent emission then can be produced.

\section{Discussion and Conclusions}
\label{con}

Via evaluation of the Lyapunov exponents of charged particles trapped in a dipole magnetic field, we found several regimes of quasi-periodic orbits with particle energies much higher than those studied with the guiding center approximation. Such orbits concentrate around the equatorial plane and the corresponding particles have a much higher perpendicular momentum than the parallel momentum, which is needed for production of strongly linearly polarized emission as observed from some FRBs. In general, one may not have an ideal dipole magnetic field. Our results suggest that as far as the magnetic field intensity varies along the magnetic field line, particles with small pitch angles will either escape or have chaotic motions.

In this paper, we didn't address the origin of such high energy particles. If the acceleration occurs in closed magnetic loops, one may consider both beta-tron and Fermi processes that  increase the perpendicular and the parallel momentum, respectively \citep{1997ApJ...485..859S}. It is interesting to note that the energy of the particle staying at rest at the saddle point is the same as that given by the Hillas criterion \citep{1984ARA&A..22..425H}. The discovery of quasi-periodic orbits at high energies may offer a mechanism to accelerate particles to the Hillas limit. The acceleration can be very fast so that a perpendicular particle distribution is established after particles with small pitch angles have already escaped or entered chaotic orbits. We therefore expect coherent strongly linearly polarized emission from such a perpendicular particle distribution.

\acknowledgments
This work is partially supported by the National Key R\&D Program of
China grant No. 2018YFA0404203 and 2018YFA0404201,
NSFC grants U1738122, U1931204, 11947404 and 11761131007,
Department of Science and Technology of Sichuan Province No. 20SYSX0294,
and by the International Partnership Program of the Chinese Academy of Sciences, grant No. 114332KYSB20170008.

\bibliography{apssamp}

\end{document}